\documentclass[10pt,conference]{IEEEtran}

\usepackage{amsmath}
\usepackage{QC,colordvi}

\usepackage{amstext,url,amssymb}

\def\QECC(#1,#2,#3,#4){[\![#1,#2,#3]\!]_{#4}}

\newcommand{\F}{{\mathbb{F}}}

\def\ket#1{\left|#1\right>}

\def\Pauli{{\cal G}_{\infty}}
\def\Cnot{{\rm ADD}}
\def\Csign{{\rm CPHASE}}

\newtheorem{theorem}{Theorem}

\newtheorem{definition}[theorem]{Definition}

\newtheorem{example}[theorem]{Example}

\begin{document}

\title{Constructions of Quantum Convolutional Codes}
 
\author{\authorblockN{Markus Grassl}
\authorblockA{
Institut f\"ur Algorithmen und Kognitive Systeme\\
Fakult\"at f\"ur Informatik, Universit\"at Karlsruhe (TH)\\
Am Fasanengarten 5, 76\,128 Karlsruhe, Germany\\
Email: grassl@ira.uka.de}
\and
\authorblockN{Martin R\"otteler}
\authorblockA{
NEC Laboratories America, Inc.\\
4 Independence Way, Suite 200\\
Princeton, NJ 08540, USA\\
Email: mroetteler@nec-labs.com
}
}

\maketitle

\pagestyle{plain}

\begin{abstract}
  We address the problems of constructing quantum convolutional codes
  (QCCs) and of encoding them. The first construction is a CSS-type
  construction which allows us to find QCCs of rate 2/4. The second
  construction yields a quantum convolutional code by applying a
  product code construction to an arbitrary classical convolutional
  code and an arbitrary quantum block code. We show that the resulting
  codes have highly structured and efficient encoders. Furthermore, we
  show that the resulting quantum circuits have finite depth,
  independent of the lengths of the input stream, and show that this
  depth is polynomial in the degree and frame size of the code.
\end{abstract}


\section{Introduction}

Similar to the classical case a quantum convolutional code (QCC)
encodes an incoming stream of quantum information into an outgoing
stream.  A basic theory of quantum convolutional codes obtained from
infinite stabilizer matrices has been developed recently, see
\cite{OlTi04}.

Only few constructions of quantum convolutional codes are known, see
\cite{Chau:98,Chau:99,AlPa04,FoGu05,FGG05,GR:2005,OlTi04}. In this
paper, we construct some new quantum convolutional codes using a
CSS-type construction which uses the same principle as the CSS
construction for block codes \cite{NC:2000}. Furthermore, we
revisit the product code construction introduced in \cite{GR:2005} and
show that for these codes the algorithm presented in \cite{GR:2006}
for computing a non-catastrophic encoder takes a particularly simple
form. This allows us to show that the depth of the encoding circuit is
polynomial in the frame size and the constraint length of the code.

\section{Quantum convolutional codes}\label{sec:qccs}
\subsection{Basic definitions}

QCCs are defined as infinite versions of quantum stabilizer codes.
The appropriate generalization of stabilizer block codes to QCCs is
provided by the polynomial formalism introduced in \cite{OlTi04}.  We
briefly sketch this approach.\footnote{We describe the approach for
  $q$ dimensional subsystems (qudits) which is a straightforward
  generalization of the binary case.}

The code is specified by its stabilizer which is a subgroup of the
infinite version $\Pauli$ of the Pauli group, which consists of tensor
products of generalized Pauli matrices acting on an semi-infinite
stream of qudits. The stabilizer can be described by a matrix with
polynomial entries
\begin{equation}\label{stab-mat}
S(D)=(X(D)|Z(D))
\in\F_q[D]^{(n-k)\times 2n}.
\end{equation}

\begin{definition}\label{def:QCCparams}
  Let $\cal C$ be a QCC defined by a full-rank stabilizer matrix as in
  eq.~(\ref{stab-mat}). Then $n$ is called the frame size, $k$ the
  number of logical qudits per frame, and $k/n$ the rate of the QCC. The
  constraint lengths are defined as $\nu_i = \max_{1 \leq j \leq
    n}(\max(\deg X_{ij}(D),\deg Z_{ij}(D)))$, the overall constraint
  length is the defined as the sum $\nu = \sum_{i=1}^{n-k} \nu_i$, and
  the memory $m$ is given by $m= \max_{1 \leq i \leq n-k} \nu_i$.
\end{definition}

Like in the classical case, a QCC can also be described in terms of a
semi-infinite stabilizer matrix $S$ which has entries in $\F_q \times
\F_q$. First, we write $S(D) = \sum_{i=0}^m G_i D^i$ using $m+1$
matrices $G_0, G_1, \ldots, G_m$ which are of size $(n-k) \times n$
each.  Then we define the semi-infinite matrix
\begin{equation}\label{semi-inf}
S:=\left(
\begin{array}{cccccccc}
G_0 & G_1 & \ldots & G_m    & 0      & \ldots &        & \\
0 &   G_0 & G_1    & \ldots & G_m    & 0      & \ldots & \\
\vdots &         & \ddots & \ddots &        & \ddots &\ddots
\end{array}
\right).
\end{equation}
Note that $S$ has a block band structure where each block is of size
$(n-k)\times (m+1) n$. A useful property of $S$ is that every qudit in
the semi-infinite stream of qudits is acted upon non-trivially by only
a finite number of generators. Moreover, those generators have bounded
support.  Hence their eigenvalues can be measured as soon as the 
corresponding qudits have been received.  Therefore, it is possible to
compute the error syndrome for the quantum convolutional code online.

There is a condition to check whether $S$ is well-defined, {\em i.\,e.}, if
it defines a commutative subgroup of $\Pauli$ \cite{OlTi04}. If $S(D)
= (X(D) | Z(D))$ as in eq.~(\ref{stab-mat}), then the condition of
symplectic orthogonality of $S$ translates to 
\begin{equation}\label{eq:symp_condition}
X(D) Z(1/D)^t - Z(D) X(1/D)^t = 0.
\end{equation}

\begin{example}
  As an example we consider the QCC defined by the stabilizer matrix
  (see \cite{FoGu05}) 
\[
S(D)=\left(
\begin{array}{ccc|ccc}
  1+D&   1& 1+D&  0  &  D  & D\\
    0&   D&   D& 1+D & 1+D & 1
\end{array}
\right).
\]
This code is derived from the classical $\F_4$-linear code generated
by $(1+D, 1+\omega D, 1+\omega^2 D)$. We can easily check
self-orthogonality by computing $X(D) Z(1/D)^t - Z(D) X(1/D)^t$ which
turns out to be the $2\times 2$ all zero matrix. Hence the code indeed
is self-orthogonal to all shifted versions of itself, {\em i.\,e.}, it
defines a QCC where $n=3$, $k=1$, and $m=1$. To illustrate the
structure in terms of Pauli matrices we consider the corresponding
semi-infinite stabilizer matrix, which is given (in Pauli form) as
follows:
\[
\arraycolsep0.4\arraycolsep
S=
\left(
\begin{array}{cccccccccccccc}
X&X&X&X&Z&Y& & & & & & & \\
Z&Z&Z&Z&Y&X& & & & & & & \\
 & & &X&X&X&X&Z&Y& & & & \\
 & & &Z&Z&Z&Z&Y&X& & & & \\
 & & & & & &X&X&X&X&Z&Y& \\
 & & & & & &Z&Z&Z&Z&Y&X& \\
 & & & & & & & & & & & & \ddots
\end{array}
\right)
\]
It is easy to see that the QCC corresponding to $S$ can correct an
arbitrary number of errors, as long as they do not occur in {\em
  bursts}, meaning in this example that at least six unaffected qubits
are between two erroneous ones.
\end{example}

\subsection{Encoding circuits}
In \cite{GRB:2003} it has been shown that for any block quantum
error-correcting code ${\cal C}=\QECC(n,k,d,q)$ there is quantum
circuit of polynomial size for encoding.  In order to encode $k$
qudits into $n$ qudits, the circuit acts on the $k$ input qudits and
$n-k$ ancillae which are initialized in the state $\ket{0}$.  The
input state can be described by a $Z$-only stabilizer matrix
$S_0=(X|Z)=(0|I\:0)$, where $I$ is an $(n-k)\times(n-k)$ identity
matrix.  The operation of the encoding circuit corresponds to a
transformation changing $S_0$ into the stabilizer $S$ of the quantum
code.  This idea can be adapted to quantum convolutional codes (see
\cite{GR:2006} for qubit codes).  The encoding circuit can be realized
by generalized Clifford gates whose action is summarized in
Table~\ref{table:clifford}, for the gates and the corresponding 
actions, see \cite[Theorem~2]{GRB:2003}.

\begin{table}[tb]\small
\caption{\small Action of various generalized Clifford operations. 
Conjugation by the unitary $U$ corresponds to the action of 
$\overline{U}$ on the columns of $S(D)=(X(D)|Z(D))$.
\label{table:clifford}}
\vskip-2ex
\[\def\arraystretch{1.5}
\begin{array}{@{}cc@{}}
\hline
\text{unitary gate $U$} & \text{matrix $\overline{U}$}\\
\hline\\[-2ex]
\text{Fourier transform ${\rm DFT}$} &
\overline{{\rm DFT}}
={\def\arraystretch{1}\arraycolsep0.5\arraycolsep\begin{pmatrix}0&-1\\1&0\end{pmatrix}}\in\F_q^{2\times 2}\\[3ex]
\hline\\[-2ex]
\text{multiplication gate $M_\gamma$}&
\overline{M_\gamma}
={\def\arraystretch{1}\arraycolsep0.5\arraycolsep\begin{pmatrix}\gamma^{-1}&0\\0&\gamma\end{pmatrix}}\in\F_q^{2\times 2}\\[3ex]
\hline\\[-2ex]
\text{diagonal gate $P_\gamma$}&
\overline{P_\gamma}
={\def\arraystretch{1}\arraycolsep0.5\arraycolsep\begin{pmatrix}1&\gamma\\0&1\end{pmatrix}}\in\F_q^{2\times 2}\\[3ex]
\hline\\[-2ex]
\Cnot^{(i,j+\ell n)}, i\not\equiv j\pmod n &
\overline{\Cnot}=
{\def\arraystretch{1}\arraycolsep0.5\arraycolsep
\left(\begin{array}{@{}cc|cc@{}}
1 & D^\ell & 0         & 0\\
0 & 1      & 0         & 0\\
\hline
0 & 0      & 1         & 0\\
0 & 0      & -D^{-\ell} & 1
\end{array}\right)}\\[5ex]
\hline\\[-2ex]
P_\ell:=\text{\Csign}^{(i,i+\ell n)},\ell\ne 0 &
\overline{P_{\ell}}=
{\def\arraystretch{1}\arraycolsep0.5\arraycolsep\begin{pmatrix}
1 & D^\ell-D^{-\ell}\\
0 & 1
\end{pmatrix}}\\[3ex]
\hline
\end{array}
\]
\vskip-4.5ex
\end{table}

\begin{figure*}[hbt]
$$
\begin{array}{r|llll|l|l}
\nu & g_1(D) & g_2(D) & g_3(D) & g_4(D) & d^\bot & N_{d^\bot}\\
\hline
 3 & 1100        & 1110        & 1001        & 1101        & 3 & 2 \\
\hline
 4 & 11001       & 11101       & 10011       & 10111	   & 4 & 1 \\
 4 & 10001       & 10101       & 11011       & 11111	   & 4 & 1 \\
\hline
 5 & 110010      & 111010      & 100001      & 110111	   & 5 & 14\\
\hline
 6 & 1010010     & 1111010     & 1000101     & 1100111	   & 6 & 63\\
\hline
 7 & 10101001    & 11111001    & 10000011    & 11000111	   & 6 & 8 \\
\hline
 8 & 101100001   & 100100101   & 111110011   & 111011011   & 6 & 2 \\
\hline
 9 & 1001001001  & 1100111101  & 1110110111  & 1010101111  & 7 & 10\\
\hline
10 & 11011011001 & 10100001101 & 10011000011 & 11001001111 & 8 & 67 \\
\hline
11 & 101101100011 & 101010110011 & 111101001011 & 110001101111 & 8 & 25
\end{array}
$$
\caption{Generators for self-orthogonal binary convolutional codes of
  rate $1/4$ yielding quantum convolutional codes of rate $2/4$ found
  by random search.\label{table:QCC_CSS_1_2}} 
\end{figure*}

\section{Efficient encoders for CSS type QCCs}\label{sec:CSS_codes}
The CSS-like construction of QCCs uses two classical convolutional
codes $C_1=(n,k_1)$ and $C_2=(n,n-k_2)$ with equal frame length $n$ and $C_2^\bot
\subseteq C_1$.  The stabilizer matrix $(X(D)|Z(D))$ is of block
diagonal form, given by
\begin{equation}\label{eq:CSS_code}
\left(
\begin{array}{c|c}
H_2(D) & 0\\
0      & H_1(D)
\end{array}
\right)\in \F_q[D]^{(n-k_1+k_2)\times 2n},
\end{equation}
where $H_1(D),H_2(D)$ denote parity check matrices of $C_1$ and $C_2$,
respectively.  We assume that both $H_1(D)$ and $H_2(D)$ correspond to
non-catastrophic, delay-free encoders and have full ranks $n-k_1$ and
$k_2$, respectively.  This implies that their Smith normal form is
$(I\: 0)$ with a suitable unit matrix $I$ (see
\cite[Chapter~2]{JoZi99}).  In particular there are unimodular
matrices $A_1(D)\in \F_q[D]^{k_2\times k_2}$ and $B_1(D)\in
\F_q^{n\times n}$ such that
\begin{equation}
A_1(D) H_2(D) B_1(D) = (I\: 0).
\end{equation}
There is an algorithm for computing the Smith normal form and the
transformation matrices $A_1(D)$ and $B_1(D)$ whose bit-complexity is
polynomial in the size and degree of the matrix $H_2(D)$ \cite{Ka85}.
This implies that the corresponding quantum circuit implementing the
matrix $B_1(D)$ can be realized using polynomially many generalized
Clifford gates.  The transformed stabilizer matrix is given by
\begin{equation}
(X'(D)|Z'(D))=\left(
\begin{array}{cc|c}
I & 0 & 0\\
0 & 0 & Z_2(D)
\end{array}
\right).
\end{equation}
The action of the gates results in a modified $Z$-part as well.  From
condition (\ref{eq:symp_condition}) it follows that $Z_2(D)$ is of the
form $Z_2(D)=(0\: Z'_2(D))$ with $Z_2'(D)\in\F_q[D]^{(n-k_1)\times
  (n-k_2)}$. Using Fourier transformation (DFT) gates on the last
$n-k_2$ qudits, the stabilizer matrix reads
\begin{equation}
(X'(D)|Z'(D))=\left(
\begin{array}{cc|c}
I & 0       & 0\\
0 & Z'_2(D) & 0
\end{array}
\right).
\end{equation}
Computing the Smith form of $Z'_2(D)$ yields matrices $A_2(D)$ and
$B_2(D)$ with $A_2(D)Z'_2(D)B_2(D)=(I\:0)$.  Another Fourier transform
on the first $n-k_1+k_2$ qudits yields an $Z$-only stabilizer matrix
of the form $(0|I 0)$.  The resulting quantum code encodes $k_1-k_2$
qudits per frame of size $n$. Overall, we obtain the following result.

\begin{theorem}
  Let ${\cal C}$ be a quantum convolutional code constructed using the
  CSS-like construction from two classical convolutional codes $C_1$
  and $C_2$ with stabilizer matrix as in
  eq.~(\ref{eq:CSS_code}). Denote the frame size with $n$ and the
  constraint length with $\nu$.  Then ${\cal C}$ has an encoding
  circuit whose depth is finite, {\em i.\,e.}, does not depend on the
  length of the input stream.  Furthermore, the depth of this circuit
  is upper bounded by $poly(n,\nu)$.
\end{theorem}

\section{CSS-type QCCs of rate $2/4$}
In \cite{FGG05}, optimal quantum convolutional codes of rate $1/3$ are
listed which are based on self-orthogonal binary convolutional codes
of rate $1/3$.  In order to construct quantum convolutional codes of
rate $2/4$, we search for self-orthogonal binary convolutional codes
$C$ of rate $1/4$ which have a dual code $C^\bot$ with high minimum
distance $d^\bot$.  Applying the CSS construction with
$C_1=C_2=C^\bot$, we then obtain a quantum convolutional code of rate
$2/4$ and minimum distance $d^\bot$.

The results of a randomized search for such codes is presented in
Table~\ref{table:QCC_CSS_1_2}.  The entries of the generator matrix
$g(D)=(g_1(D),g_2(D),g_3(D),g_4(D))$ of the code $C$ are given in
abbreviated form, listing the coefficients in increasing order.  For
example, the generator matrix of the first code with constraint length
$\nu=3$ is $g(D)=(1+D,1+D+D^2,1+D^3,1+D+D^3)$.  The last column lists
the number $N_{d^\bot}$ of sequences of minimum weight. Note that it
is desirable to have as few sequences of minimum weight as
possible. The size of the search space grows with $O(2^{4\nu}$), so we
have only performed an exhaustive search up to constraint length
$\nu=6$, and a randomized search for larger values of $\nu$.

\section{Efficient encoders for product codes}
\subsection{Product code construction}
The following theorem, taken from \cite{GR:2005}, allows to construct
a quantum convolutional code using a classical convolutional code and
a quantum code.
\begin{theorem}\label{th:product}
Let $C_1=(n_1,k_1)_p$ be a classical convolutional code over $\F_p$
with dual distance $d_1^\bot$ and let $G_1(D)$ be a generator matrix
of $C_1$ corresponding to a non-catastrophic, delay-free encoder.
Furthermore, let ${\cal C}$ be a quantum error-correcting code for
$q$-dimensional quantum systems ($q=p^\ell$) with minimum distance
$d_2$ and stabilizer matrix $S_2=(X|Z)$ if ${\cal C}$ is a block code
or $S_2=(X(D)|Z(D))$ if ${\cal C}$ is a convolutional code.  Then the
stabilizer matrix
\begin{equation}\label{eq:tensorprod}
G(D)=G_1(D) \otimes_p S_2
\end{equation}
defines a quantum convolutional code with minimum distance $d\le
\min(d_1^\bot,d_2)$. 
\end{theorem}

The tensor product $\otimes_p$ corresponds to the Kronecker product of
the stabilizer matrices.  We use the index $p$ to stress that the
coefficients of the polynomials in the matrix $G_1(D)$ are in the
prime field $\F_p$ while the stabilizer matrix $S_2$ might be defined
over an extension field $\F_q=\F_{p^\ell}$.

\subsection{Encoding product codes}
Instead of applying the general algorithm of \cite{GR:2006} to the
matrix $G(D)$ in order to compute an encoding circuit for the product
code, we will exploit the additional structure of the stabilizer
matrix.  The first step is to compute an inverse encoding circuit for
the quantum code ${\cal C}$ with stabilizer $S_2$.  The quantum
circuit corresponds to a symplectic transformation yielding the
trivial $Z$-only stabilizer $S_0=(0|I\:0)$. Note that the trivial
stabilizer is of this form, regardless whether the code ${\cal C}$ is
a block or a convolutional quantum code. Omitting the final Fourier
transformation gates in the quantum circuit, we obtain an $X$-only
stabilizer $S'_0=(I\:0|0)$.

Expanding the matrix $G_1(D)$ as semi-infinite matrix, we get the
following semi-infinite version of the stabilizer matrix $G(D)$ of
eq. (\ref{eq:tensorprod}):
\begin{small}
\[
\arraycolsep0.2\arraycolsep
\left(
\begin{array}{ccccccccc}
g_{11}S_2 & g_{12}S_2 & \ldots &g_{1,n_1} S_2\\
g_{21}S_2 & g_{22}S_2 & \ldots &g_{2,n_1} S_2\\
\vdots &\vdots&\ddots&\vdots\\
g_{k_1,1}S_2 & g_{k_1,2}S_2 & \ldots & g_{k_1,n_1} S_2\\[2ex]
&&&g_{11}S_2 & g_{12}S_2 & \ldots &g_{1,n_1} S_2\\
&&&g_{21}S_2 & g_{22}S_2 & \ldots &g_{2,n_1} S_2\\
&&&\vdots &\vdots&\ddots&\vdots\\
&&&g_{k_1,1}S_2 & g_{k_1,2}S_2 & \ldots & g_{k_1,n_1} S_2\\
&&&&&&\vdots&\ddots
\end{array}
\right)
\]
\end{small}%
This matrix indicates that we have to apply the inverse encoding
circuit of the code ${\cal C}$ to every block of qudits corresponding
to the submatrices $g_{ij}S_2$.  This first step corresponds to the
leftmost boxes marked ${\rm BC}$ in the example of
Fig.~\ref{fig:circ_prod}.  The stabilizer matrix is now of the form
\begin{equation}\label{eq:CSS_prod}
G'(D)=G_1(D)\otimes_p (I\:0|0)=(G_1(D)\otimes I | 0).
\end{equation}
This $X$-only generator matrix corresponds to a CSS code (see
eq.~(\ref{eq:CSS_code})) where
\[
H_2(D)=\left(
\begin{array}{ccc}
G_1(D)\\
& \ddots\\
&& G_1(D)
\end{array}
\right).
\]
Using the algorithm of Sect.~\ref{sec:CSS_codes}, we obtain an inverse
encoding circuit for the convolutional CSS code corresponding to
$G_1(D)$.  This circuit has to be repeated $r$ times if the identity
matrix in eq.~(\ref{eq:CSS_prod}) has rank $r$.  The $j$-th copy of this
quantum circuits acts on qudits $j,j+r,j+2r,\ldots$ (see the blocks
marked ${\rm CC}_j$ in the example of Fig.~\ref{fig:circ_prod}).
Overall, we obtain the following result.

\begin{theorem}
  Let ${\cal C}$ be a quantum convolutional code which has been
  constructed using the product code construction described in
  Theorem~\ref{th:product}. Denote the frame size with $n$ and the
  constraint length with $\nu$. Then ${\cal C}$ has an encoding
  circuit whose depth is finite, {\em i.\,e.}, does not depend on the length
  of the input stream. Furthermore, the depth of this circuit is upper
  bounded by $poly(n,\nu)$.
\end{theorem}

\subsection{QCCs from products of cyclic codes}
In \cite[Theorem 8]{GR:2005}, we have shown that product codes based
on Reed-Solomon codes achieve the upper bound on the minimum distance
of the resulting quantum code.  Here we consider the following
variant:
\begin{theorem}\label{theorem:cyclic}
Let $C$ be a cyclic code over $\F_q$ of composite length $n=n_1 n_2$
with $n_2|(q-1)$.  Furthermore, we assume that $C$ can be decomposed
as $C=C_1\otimes C_2$ where $C_2$ has generator polynomial
$g_2(X)=\prod_{i=1}^{d-1}(X-\alpha^i)$ where $d-1\le n_2/2$ and
$\alpha$ is an $n_2$-th root of unity in $\F_q$.  Then the code $C$ is
self-orthogonal and has dual distance $d^\bot \ge\min(\delta_1,d)$
where $\delta_1$ is the BCH bound of $C_1^\bot$.
\end{theorem}
\begin{proof}
The code $C_2$ is generated by
$$
G_2=\left(
\begin{array}{ccccc}
\alpha^0 & \alpha & \alpha^2 & \ldots & \alpha^{n_2-1}\\
\alpha^0 & \alpha^2 & \alpha^4 & \ldots & \alpha^{2(n_2-1)}\\
\vdots   & \vdots   & \vdots & \ddots & \vdots\\
\alpha^0 & \alpha^{d-1} & \alpha^{2(d-1)} & \ldots & \alpha^{(d-1)(n_2-1)}
\end{array}
\right).
$$
The inner product of row $i$ and row $j$ of $G_2$ is
$$
\sum_{\ell=0}^{n_2-1}\alpha^{(i+j)\ell}=0
$$
as $i+j\not\equiv 0 \bmod n_2$. Hence $C_2$ is self-orthogonal and so
is $C$.  The bound on the minimum distance follows from the
two-dimensional BCH bound \cite[p. 320]{Bla03}.
\end{proof}

Starting with a generator matrix $G=G_1\otimes G_2$ of a (permuted)
cyclic code as in Theorem~\ref{theorem:cyclic}, we can construct
convolutional quantum codes of CSS type.  The semi-infinite generator
matrix of the corresponding self-orthogonal convolutional code is
formed by the copies of the generator matrix $G$ which overlap in $\mu
n_2$ positions. For $\mu=2$ we get

\begin{small}
\begin{equation}\label{eq:QCC_cyclic}
\arraycolsep0.2\arraycolsep
\left(
\begin{array}{ccccccccc}
g_{11}G_2 & g_{12}G_2 & \ldots &g_{1,n_1} G_2\\
g_{21}G_2 & g_{22}G_2 & \ldots &g_{2,n_1} G_2\\
\vdots &\vdots&\ddots&\vdots\\
g_{k_1,1}G_2 & g_{k_1,2}G_2 & \ldots & g_{k_1,n_1} G_2\\[2ex]
&&g_{11}G_2 & g_{12}G_2 & \ldots &g_{1,n_1} G_2\\
&&g_{21}G_2 & g_{22}G_2 & \ldots &g_{2,n_1} G_2\\
&&\vdots &\vdots&\ddots&\vdots\\
&&g_{k_1,1}G_2 & g_{k_1,2}G_2 & \ldots & g_{k_1,n_1} G_2\\
&&&&&\vdots&\ddots
\end{array}
\right).
\end{equation}
\end{small}%
The inner product of any two rows of this matrix is zero, as already
$G_2\cdot G_2^t=0$.  The dual distance of the dual of the convolution
code defined by eq.~(\ref{eq:QCC_cyclic}) is lower bounded by the dual
distance of $C$, as any sequence in the dual of the convolutional code
fulfills the parity checks given by the matrix $G$.  Note that the
encoder corresponding to the matrix (\ref{eq:QCC_cyclic}) might be
catastrophic. Then, in some cases, the minimal non-catastrophic
encoder can have constraint length zero, {\em i.\,e.}, corresponds to
a block code.

\section{Example}
We illustrate the product construction and the corresponding encoding
circuit using the five qubit code $\QECC(5,1,3,2)$ and a classical
convolutional code of rate $R=2/3$.
\begin{figure}[hbt]
\unitlength0.65\unitlength
\centerline{
\inputwires[,,$\left.\ket{\phi_{\text{out}}}\rule{0pt}{45\unitlength}\right\{\!\!$](5)
\kern-15\unitlength%
\CNOT(1,4,5)%
\kern-4\unitlength%
\rlap{\OneQubitGate[3](2,2){$H$}}%
\rlap{\OneQubitGate[2](1,1){$H$}}%
\OneQubitGate(1,2){$H$}%
\kern-4\unitlength%
\rlap{\CNOT(1,2,5)}%
\rlap{\CNOT(1,3,5)}%
\CNOT(1,4,5)\kern-5\unitlength%
\kern-4\unitlength%
\rlap{\CNOT(2,3,5)}%
\rlap{\CNOT(2,4,5)}%
\CNOT(2,5,5)%
\kern-4\unitlength%
\OneQubitGate(5,5){$H$}%
\kern-4\unitlength%
\CNOT(2,5,5)\kern-5\unitlength%
\kern-4\unitlength%
\CNOT(3,4,5)%
\rlap{\OneQubitGate[1](4,4){$H$}}%
\OneQubitGate(1,1){$H$}%
\kern-4\unitlength%
\rlap{\CNOT(3,4,5)}%
\CNOT(3,5,5)
\kern-4\unitlength%
\OneQubitGate(4,5){$H$}%
\rlap{\OneQubitGate[4](1,1){$H$}}%
\rlap{\OneQubitGate[3](1,1){$H$}}%
\rlap{\OneQubitGate[2](1,1){$H$}}%
\OneQubitGate(1,2){$H$}%
\kern-7\unitlength%
\outputwires[$\ket{0}$,$\ket{0}$,$\ket{0}$,$\ket{0}$,$\ket{\phi_{\text{in}}}$](5)
\kern-13\unitlength}
\caption{Inverse Encoding Circuit for the five qubit code.\label{fig:circ5qubit}}
\end{figure}

Using the inverse encoding circuit shown in Fig.~\ref{fig:circ5qubit},
the stabilizer
\begin{eqnarray}
S&=&
\left(
\begin{array}{*5c|*5c}
1&0&0&1&0&1&1&1&1&0\\
0&1&0&0&1&0&1&1&1&1\\
1&0&1&0&0&1&0&1&1&1\\
0&1&0&1&0&1&1&0&1&1
\end{array}
\right)\label{eq:stab5qubit}
\end{eqnarray}
of the five qubit code is transformed into
\begin{eqnarray*}
S'&=&
\left(
\begin{array}{*5c|*5c}
0&0&0&0&0 &1&0&0&0&0\\
0&0&0&0&0 &0&1&0&0&0\\
0&0&0&0&0 &1&1&1&0&0\\
0&0&0&0&0 &0&1&1&1&0
\end{array}
\right).
\end{eqnarray*}
Note that the two Hadamard transforms on the fourth qubit cancel, but
when omitting the final four Hadamard transformations, we obtain an
$X$-only stabilizer.

In \cite[Table 8.14]{JoZi99} we find a nonsystematic rate
$R=2/3$ convolutional code with memory $\nu=2$ and free distance $d_{\text{free}}=3$.
An encoding matrix for that code is
\begin{eqnarray*}
G(D)&=&
\left(
\begin{array}{*3c}
  D+D^2 &   1     &  1+D^2\\
   1    & D + D^2 &  1+D+D^2  
\end{array}
\right).
\end{eqnarray*}
A minimal polynomial generator matrix for the dual code is given by
\begin{eqnarray}
H(D)&=&
\left(
\begin{array}{c}
 1 + D + D^4\\
  1 + D^2 + D^3 + D^4\\
 1 + D^2 + D^4
\end{array}
\right)^t.\label{eq:conv_code}
\end{eqnarray}

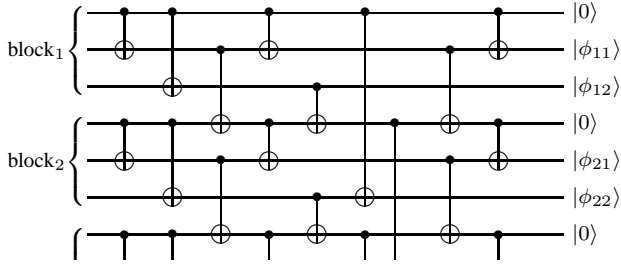
\begin{figure}[hbt]
\unitlength0,7\unitlength
\centerline{\footnotesize%
\inputwires[,$\left.\rule{0pt}{28\unitlength}\text{block}_1\right\{\!\!$,,,
$\left.\rule{0pt}{28\unitlength}\text{block}_2\right\{\!\!$,,,
$\left.\rule{0pt}{28\unitlength}\right\{\!\!$](7)%
\kern-15\unitlength%
\rlap{\CNOT(1,2,7)}%
\rlap{\CNOT(1,2,4)}%
\CNOT(1,2,1)%
\kern-4\unitlength%
\rlap{\CNOT(1,3,7)}%
\rlap{\CNOT(1,3,4)}%
\CNOT(1,3,1)%
\kern-4\unitlength%
\rlap{\CNOT(2,4,7)}%
\rlap{\CNOT(2,4,4)}%
\CNOT(2,4,1)%
\kern-4\unitlength%
\rlap{\CNOT(1,2,7)}%
\rlap{\CNOT(1,2,4)}%
\CNOT(1,2,1)%
\kern-4\unitlength%
\rlap{\CNOT(3,4,7)}%
\rlap{\CNOT(3,4,4)}%
\CNOT(3,4,1)%
\kern-4\unitlength%
\rlap{\CNOT(1,6,7)}%
\CNOT(1,6,1)\kern-10\unitlength%
\kern-4\unitlength%
\CNOT(4,9,7)%
\rlap{\CNOT(2,4,7)}%
\rlap{\CNOT(2,4,4)}%
\CNOT(2,4,1)%
\kern-4\unitlength%
\rlap{\CNOT(1,2,7)}%
\rlap{\CNOT(1,2,4)}%
\CNOT(1,2,1)%
\outputwires[$\ket{0}$,$\ket{\phi_{11}}$,$\ket{\phi_{12}}$,$\ket{0}$,$\ket{\phi_{21}}$,$\ket{\phi_{22}}$,$\ket{0}$](7)
\begin{picture}(0,0)
\put(0,-4){\makebox(0,0)[tr]{\White{\rule{300\unitlength}{100\unitlength}}}}
\end{picture}
}
\caption{Quantum circuit transforming the stabilizer
  $(X(D)|Z(D)):=(H(D)|0)$ into the simple form $(XII)$.\label{fig:class_circ}}
\end{figure}

If we apply our algorithm to the stabilizer code with $X$-only
stabilizer matrix $(X(D)|Z(D)):=(H(D)|0)$ we obtain the circuit shown
in Fig.~\ref{fig:class_circ}.  The resulting transformed stabilizer is
of the simple form $(XII)$

Using the product construction, we take the tensor product of the
stabilizer matrix $S$ in eq.~(\ref{eq:stab5qubit}) and the generator
matrix $H(D)$ of the binary convolutional code in
eq.~(\ref{eq:conv_code}).  The stabilizer matrix has the form
\begin{equation}\label{eq:prod1}
S_{\text{product}}=\bigl( H(D)\otimes S_X\bigm| H(D)\otimes S_Z\bigr),
\end{equation}
where $S_X$ and $S_Z$ denote the corresponding parts of $S$.

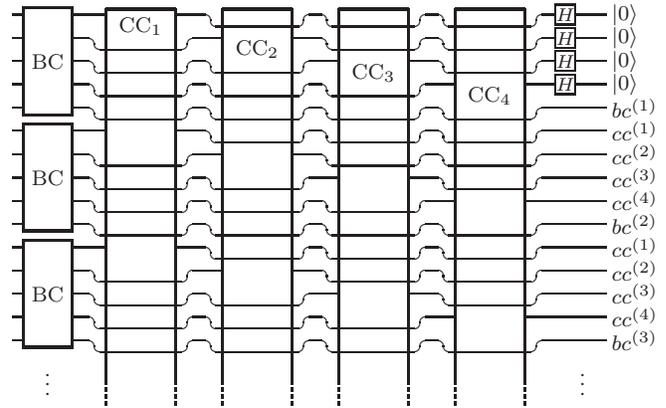
\begin{figure}[hbt]
\def\runter{%
\begin{picture}(20,10)
\put(0,0){\line(1,0){5}}
\put(5,-5){\oval(10,10)[tr]}
\put(15,-5){\oval(10,10)[bl]}
\put(15,-10){\line(1,0){5}}
\end{picture}%
}
\def\rauf{%
\begin{picture}(20,10)
\put(0,0){\line(1,0){5}}
\put(5,5){\oval(10,10)[br]}
\put(15,5){\oval(10,10)[tl]}
\put(15,10){\line(1,0){5}}
\end{picture}%
}

\unitlength0.44\unitlength\footnotesize%
\begin{picture}(60,300)
\multiput(0,10)(0,20){15}{\line(1,0){10}}
\thicklines
\multiput(10,5)(0,100){3}{\framebox(40,90){${\rm BC}$}}
\thinlines
\multiput(50,10)(0,20){15}{\line(1,0){10}}
\multiput(30,-20)(0,-8){3}{\makebox(0,0){.}}
\end{picture}%
\begin{picture}(100,300)
\multiput(0,10)(0,100){3}{\runter}
\multiput(20,0)(0,100){3}{\line(1,0){60}}
\multiput(80,0)(0,100){3}{\rauf}
\multiput(0,30)(0,100){3}{\runter}
\multiput(20,20)(0,100){3}{\line(1,0){60}}
\multiput(80,20)(0,100){3}{\rauf}
\multiput(0,50)(0,100){3}{\runter}
\multiput(20,40)(0,100){3}{\line(1,0){60}}
\multiput(80,40)(0,100){3}{\rauf}
\multiput(0,70)(0,100){3}{\runter}
\multiput(20,60)(0,100){3}{\line(1,0){60}}
\multiput(80,60)(0,100){3}{\rauf}
\multiput(0,90)(0,100){3}{\line(1,0){20}}
\multiput(80,90)(0,100){3}{\line(1,0){20}}
\thicklines
\put(20,-25){\line(0,1){320}}
\multiput(20,-25)(0,-6){4}{\line(0,-1){3}}
\put(20,295){\line(1,0){60}}
\put(50,280){\makebox(0,0){${\rm CC}_1$}}
\put(80,-25){\line(0,1){320}}
\multiput(80,-25)(0,-6){4}{\line(0,-1){3}}
\end{picture}%
\begin{picture}(100,300)
\multiput(0,10)(0,100){3}{\runter}
\multiput(20,0)(0,100){3}{\line(1,0){60}}
\multiput(80,0)(0,100){3}{\rauf}
\multiput(0,30)(0,100){3}{\runter}
\multiput(20,20)(0,100){3}{\line(1,0){60}}
\multiput(80,20)(0,100){3}{\rauf}
\multiput(0,50)(0,100){3}{\runter}
\multiput(20,40)(0,100){3}{\line(1,0){60}}
\multiput(80,40)(0,100){3}{\rauf}
\multiput(0,70)(0,100){3}{\line(1,0){20}}
\multiput(80,70)(0,100){3}{\line(1,0){20}}
\multiput(0,90)(0,100){3}{\runter}
\multiput(20,80)(0,100){3}{\line(1,0){60}}
\multiput(80,80)(0,100){3}{\rauf}
\thicklines
\put(20,-25){\line(0,1){320}}
\multiput(20,-25)(0,-6){4}{\line(0,-1){3}}
\put(20,295){\line(1,0){60}}
\put(50,260){\makebox(0,0){${\rm CC}_2$}}
\put(80,-25){\line(0,1){320}}
\multiput(80,-25)(0,-6){4}{\line(0,-1){3}}
\end{picture}%
\begin{picture}(100,300)
\multiput(0,10)(0,100){3}{\runter}
\multiput(20,0)(0,100){3}{\line(1,0){60}}
\multiput(80,0)(0,100){3}{\rauf}
\multiput(0,30)(0,100){3}{\runter}
\multiput(20,20)(0,100){3}{\line(1,0){60}}
\multiput(80,20)(0,100){3}{\rauf}
\multiput(0,50)(0,100){3}{\line(1,0){20}}
\multiput(80,50)(0,100){3}{\line(1,0){20}}
\multiput(0,70)(0,100){3}{\runter}
\multiput(20,60)(0,100){3}{\line(1,0){60}}
\multiput(80,60)(0,100){3}{\rauf}
\multiput(0,90)(0,100){3}{\runter}
\multiput(20,80)(0,100){3}{\line(1,0){60}}
\multiput(80,80)(0,100){3}{\rauf}
\thicklines
\put(20,-25){\line(0,1){320}}
\multiput(20,-25)(0,-6){4}{\line(0,-1){3}}
\put(20,295){\line(1,0){60}}
\put(50,240){\makebox(0,0){${\rm CC}_3$}}
\put(80,-25){\line(0,1){320}}
\multiput(80,-25)(0,-6){4}{\line(0,-1){3}}
\end{picture}%
\begin{picture}(100,300)
\multiput(0,10)(0,100){3}{\runter}
\multiput(20,0)(0,100){3}{\line(1,0){60}}
\multiput(80,0)(0,100){3}{\rauf}
\multiput(0,30)(0,100){3}{\line(1,0){20}}
\multiput(80,30)(0,100){3}{\line(1,0){20}}
\multiput(0,50)(0,100){3}{\runter}
\multiput(20,40)(0,100){3}{\line(1,0){60}}
\multiput(80,40)(0,100){3}{\rauf}
\multiput(0,70)(0,100){3}{\runter}
\multiput(20,60)(0,100){3}{\line(1,0){60}}
\multiput(80,60)(0,100){3}{\rauf}
\multiput(0,90)(0,100){3}{\runter}
\multiput(20,80)(0,100){3}{\line(1,0){60}}
\multiput(80,80)(0,100){3}{\rauf}
\thicklines
\put(20,-25){\line(0,1){320}}
\multiput(20,-25)(0,-6){4}{\line(0,-1){3}}
\put(20,295){\line(1,0){60}}
\put(50,220){\makebox(0,0){${\rm CC}_4$}}
\put(80,-25){\line(0,1){320}}
\multiput(80,-25)(0,-6){4}{\line(0,-1){3}}
\multiput(130,-20)(0,-8){3}{\makebox(0,0){.}}
\end{picture}%
\rlap{\OneQubitGate[14](1,1){\scriptsize$H$}}%
\rlap{\OneQubitGate[13](1,1){\scriptsize$H$}}%
\rlap{\OneQubitGate[12](1,1){\scriptsize$H$}}%
\OneQubitGate(1,12){\scriptsize$H$}%
\outputwires[$\ket{0}$,$\ket{0}$,$\ket{0}$,$\ket{0}$,%
$bc^{(1)}$,$cc^{(1)}$,$cc^{(2)}$,$cc^{(3)}$,$cc^{(4)}$,%
$bc^{(2)}$,$cc^{(1)}$,$cc^{(2)}$,$cc^{(3)}$,$cc^{(4)}$,%
$bc^{(3)}$](15)%
\vskip30\unitlength

\caption{Schematic inverse encoding circuit for the quantum
  convolutional code of rate $R=11/15$ obtained by the product code
  from the quantum block code $\QECC(5,1,3,2)$ and a classical
  convolutional code with rate $R=2/3$.
\label{fig:circ_prod}}
\end{figure}

Note that the circuit shown in Fig.~\ref{fig:circ5qubit} corresponds
to a binary symplectic matrix $T=T_1 T_2$, i.e., $S T=S'$, where $T_2$
corresponds to the last four Hadamard gates.  Replicating the circuit
without these Hadamard gates three times as indicated in
Fig.~\ref{fig:circ_prod}, we get the matrix $I_3\otimes T_1$, where
$I_3$ denotes a $3\times 3$ identity matrix.  Now the $Z$-part of the
stabilizer is zero, and the $X$-part has the form
\[
\left(
\begin{array}{c}
 1 + D + D^4\\
  1 + D^2 + D^3 + D^4\\
 1 + D^2 + D^4
\end{array}
\right)^t 
\otimes
\left(
\begin{array}{*5c}
1&0&0&0&0\\
0&1&0&0&0\\
1&1&1&0&0\\
0&1&1&1&0
\end{array}
\right).
\]
So replicating the circuit of Fig.~\ref{fig:class_circ} four times
(but spread out to any every fifth qubit), we get an $X$-only stabilizer.  The
final four Hadamard gates in Fig.~\ref{fig:circ_prod} transform it
into a $Z$-only stabilizer.

The structure of the whole encoding circuit is illustrated in
Fig.~\ref{fig:circ_prod}. Only the first block is shown, but every
quantum gate in the circuit has to be applied repeatedly, shifted by
the corresponding number of qubits.  Each block encodes 11 qubits into
15. The inputs marked with $bc^{(i)}$ correspond to the input of the
$i$-th copy of the block code $[\![5,1,3]\!]$, the inputs of the four
copies of the convolutional code are marked with $cc^{(j)}$.  The
boxes marked with ${\rm BC}$ correspond to the encoder for the block
code in Fig.~\ref{fig:circ5qubit}, the blocks ${\rm CC}_j$ correspond
to the encoding circuit for the convolutional code in
Fig.~\ref{fig:class_circ}.

\enlargethispage{-17mm}

\section{Conclusions}

The problem of constructing quantum convolutional codes and their
encoders was addressed. Using a CSS-type construction, we derived new
examples of QCCs of rate $2/4$. For constraint lengths up to $\nu=6$ we
performed an exhaustive search of the search space, and for constraint
lengths up to $11$ we employed a randomized search which found several
good codes. Using a product code construction which takes as inputs a
classical convolutional code on the one hand and a quantum block code
on the other, it is possible to derive many examples of QCCs. We show
that these codes all have the property that their encoder is of
polynomial depth. We conjecture that any stabilizer QCC has a
polynomial depth encoder. It seems that a more detailed study of the
algorithm given in \cite{GR:2006}, which is based on iterative Smith
normal form computation on the stabilizer matrix, would be required to
resolve this question.

\end{document}